\documentclass{aipproc}
\layoutstyle{8x11double}

\def\gtwid{\mathrel{\raise.3ex\hbox{$>$\kern-.75em\lower1ex\hbox{$\sim$}}}}
\def\ltwid{\mathrel{\raise.3ex\hbox{$<$\kern-.75em\lower1ex\hbox{$\sim$}}}}

\begin{document}

\title{Axions~05}


\author{Pierre Sikivie}{
  address={Institute for Fundamental Theory\\
           Physics Department, University of Florida\\ 
           Gainesville, FL 32611-8440, USA}}

\begin{abstract}

The Strong CP Problem and its resolution through the existence of an 
axion are briefly reviewed.  The combined constraints from accelerator
searches, the evolution of red giants and the duration of the SN 1987a 
neutrino pulse require the axion mass $m_a$ to be less than $3 \cdot
10^{-3}$ eV.  On the other hand, the requirement that the axion does 
not overclose the universe implies a lower bound on $m_a$ of order
$10^{-6}$ eV.  This lower bound can, however, be relaxed in a number 
of ways.  If $m_a$ is near the lower bound, axions are an important 
contribution to the energy density of the universe in the form of 
(very) cold dark matter.  Dark matter axions can be searched for on 
Earth by stimulating their conversion to microwave photons in an
electromagnetic cavity permeated by a magnetic field.  Using this 
technique, limits on the local halo density have been placed by the 
Axion Dark Matter eXperiment.  I'll give a status report on ADMX 
and its upgrade presently under construction.  I'll also report on 
recent results from solar axion searches and laser experiments.

\end{abstract}

\maketitle


\section{Introduction}

The standard model of elementary particles has been an extraordinary 
breakthrough in our description of the physical world.  It agrees
with experiment and observation disconcertingly well and surely 
provides the foundation for all further progress in our field.  It 
does however present us with a puzzle.

Indeed the action density includes, in general, a term
\begin{equation}
{\cal L}_{\rm stand~mod} = ... 
~+~{\theta g^2\over 32\pi^2} G^a_{\mu\nu} \tilde G^{a\mu\nu}
\label{ggdual}
\end{equation}
where $G^a_{\mu\nu}$ are the QCD field strengths, $g$ is the QCD 
coupling constant and $\theta$ is a parameter.  The dots represent 
all the other terms in the action density, i.e. the terms that lead 
to the numerous successes of the standard model.  Eq.~(\ref{ggdual})
perversely shows the one term that isn't a success.  Using the 
statement of the chiral anomaly \cite{abj}, one can show three 
things about that term.  First, that QCD physics depends on the 
value of the parameter $\theta$ because in the absence of such 
dependence QCD would have a $U_A(1)$ symmetry in the chiral limit, 
and we know QCD has no such $U_A(1)$ symmetry \cite{SW}.  Second, 
that $\theta$ is cyclic, that is to say that physics at $\theta$ 
is indistinguishable from physics at $\theta + 2\pi$.  Third, that 
an overall phase in the quark mass matrix $m_q$ can be removed by 
a redefinition of the quark fields only if, at the same time, 
$\theta$ is shifted to $\theta - \arg\det m_q$.  The combination of 
standard model parameters $\bar{\theta} \equiv \theta - \arg \det m_q$ 
is independent of quark field redefinitions.  Physics therefore 
depends on $\theta$ solely through $\bar{\theta}$.

Since physics depends on $\bar{\theta}$, the value of $\bar{\theta}$ 
is determined by experiment.  The term shown in Eq.~(\ref{ggdual}) 
violates P and CP.  This source of P and CP violation is incompatible 
with the experimental upper bound on the neutron electic dipole moment
\cite{ned} unless $|\bar{\theta}| < 10^{-10}$.  The puzzle aforementioned
is why the value of $\bar{\theta}$ is so small.  It is usually referred 
to as the ``Strong CP Problem".  If there were only strong interactions, 
a zero value of $\bar{\theta}$ could simply be a consequence of P 
and CP conservation.  That would not be much of a puzzle. But there 
are also weak interactions and they, and therefore the standard model 
as a whole, violate P and CP.  So these symmetries can not be invoked 
to set $\bar{\theta} = 0$.  More pointedly, P and CP violation are
introduced in the standard model by letting the elements of the quark 
mass matrix $m_q$ be arbitrary complex numbers \cite{KM}.  In that 
case, one expects $\arg \det m_q$, and hence $\bar{\theta}$, to be 
a random angle.  

The puzzle is removed if the action density is instead
\begin{eqnarray}
{\cal L}_{\rm stand~mod~+~axion} &=&~...
~+~{1 \over 2}\partial_\mu a \partial^\mu a\nonumber\\
&+&{g^2\over 32\pi^2}~{a(x) \over f_a}~G^a_{\mu\nu} \tilde G^{a\mu\nu}
\label{ax}
\end{eqnarray}
where $a(x)$ is a new scalar field, and the dots represent the other 
terms of the standard model.  $f_a$ is a constant with dimension of
energy.  The $a G \cdot \tilde G$ interaction in Eq.~(\ref{ax}) is 
not renormalizable.  However, there is a recipe for constructing 
renormalizable theories whose low energy effective action density 
is of the form of Eq.~(\ref{ax}).  The recipe is as follows: construct 
the theory in such a way that it has a $U(1)$ symmetry which (1) is a 
global symmetry of the classical action density, (2) is broken by the
color anomaly, and (3) is spontaneously broken.  Such a symmetry is 
called Peccei-Quinn symmetry after its inventors \cite{PQ}.  Weinberg 
and Wilczek \cite{WW} pointed out that a theory with a $U_{\rm PQ}(1)$
symmetry has a light pseudo-scalar particle, called the axion.  The 
axion field is $a(x)$.  $f_a$ is of order the expectation value that 
breaks $U_{\rm PQ}(1)$, and is called the ``axion decay constant".

In the theory defined by Eq.~(\ref{ax}), $\bar{\theta} = {a(x) \over f_a}
- \det\arg m_q$ depends on the expectation value of $a(x)$.  That
expectation value minimizes the effective potential.  The strong 
CP problem is then solved because the minimum of the QCD effective
potential $V(\bar{\theta})$ occurs at $\bar{\theta} = 0$ \cite{VW}.  
The weak interactions induce a small value for $\bar{\theta}$, of 
order $10^{-17}$ \cite{GR}, but this is consistent with experiment.  

The notion of Peccei-Quinn (PQ) symmetry may seem contrived.  Why 
should there be a $U(1)$ symmetry which is broken at the quantum 
level but which is exact at the classical level?  One should keep 
in mind, however, that the reasons for PQ symmetry may be deeper 
than we know at present.  String theory contains many examples of
symmetries which are exact classically but which are broken by 
anomalies, including PQ symmetry.  Also, within field theory, 
there are examples of theories with {\it automatic} PQ symmetry,
i.e. where PQ symmetry is a consequence of just the particle 
content of the theory without adjustment of parameters to special
values. 

The first axion models had $f_a$ of order the weak interaction
scale and it was thought that this was an unavoidable property of
axion models.  However, our conference chairman J.E. Kim and others
pointed out \cite{KSVZ,DFSZ} that the value of $f_a$ is really 
arbitrary, that it is possible to construct axion models with 
any value of $f_a$.  J.E. Kim also pointed out \cite{Kim} that 
a value of $f_a$ far from any previously known scale need not 
lead to a hierarchy problem because PQ symmetry can be broken 
by the condensates of a new technicolor-like interaction.

The properties of the axion can be derived using the methods of
current algebra \cite{curr}.  The axion mass is given in terms of 
$f_a$ by
\begin{equation}
m_a\simeq 6~eV~{10^6 GeV\over f_a}\, .
\label{ma}
\end{equation}
All the axion couplings are inversely proportional to $f_a$.
For example, the axion coupling to two photons is:
\begin{equation}
{\cal L}_{a\gamma\gamma} = -g_\gamma {\alpha\over \pi} {a(x)\over f_a}
\vec E \cdot\vec B~~~\ .
\label{aEB}
\end{equation}
Here $\vec E$ and $\vec B$ are the electric and magnetic fields,
$\alpha$ is the fine structure constant, and $g_\gamma$ is a
model-dependent coefficient of order one.  $g_\gamma=0.36$ in 
the DFSZ model \cite{DFSZ} whereas $g_\gamma=-0.97$ in the KSVZ 
model \cite{KSVZ}.  The coupling of the axion to a spin 1/2 fermion 
$f$ has the form:
\begin{equation}
{\cal L}_{a \overline f f} = i g_f {m_f \over f_a} 
a \overline f \gamma_5 f
\label{cf}
\end{equation}
where $g_f$ is a model-dependent coefficient of order one.  In the
KSVZ model the coupling to electrons is zero at tree level.  Models
with this property are called 'hadronic'.

\begin{figure}[t]
\includegraphics[height=.4\textheight]{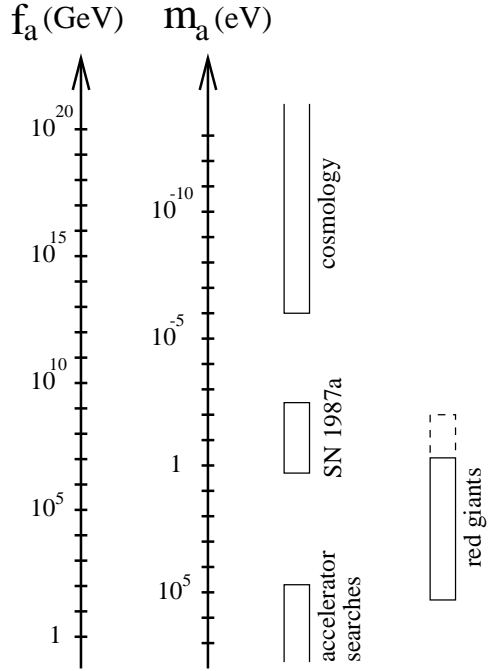}
\caption{Ranges of axion mass $m_a$, or equivalently axion
decay constant $f_a$, which have been ruled out by accelerator
searches, the evolution of red giants, the supernova SN1987a,
and finally the axion cosmological energy density.}
\label{fig1}
\end{figure}

The axion has been searched for in many places but not found \cite{arev}.
Fig.~\ref{fig1} summarizes the constraints.  Axion masses larger than 
about 50 keV are ruled out by particle physics experiments (beam dumps 
and rare decays) and nuclear physics experiments.  The next range of
axion masses, in decreasing order, is ruled out by stellar evolution 
arguments.  The longevity of red giants rules out 
200 keV $> m_a >$ 0.5 eV \cite{Dicus,Raff87} in the case of hadronic
axions, and 200 keV $> m_a > 10^{-2}$ eV \cite{Schramm} in the case of
axions with a large coupling to electrons [$g_e = 0(1)$ in Eq. \ref{cf}]. 
The duration of the neutrino pulse from supernova 1987a rules out 
2 eV $> m_a > 3 \cdot 10^{-3}$ eV \cite{1987a}.  Finally, there is a 
lower limit, $m_a \gtwid 10^{-6}$ eV, from cosmology which will be 
discussed in detail in the next section.  This leaves open an ``axion
window": $3 \cdot 10^{-3} > m_a \gtwid 10^{-6}$ eV.  Note however that 
the lower edge of this window ($10^{-6}$ eV) is much softer than its
upper edge.

\section{Axion cosmology}

The implications of the existence of an axion for the history of the
early universe may be briefly described as follows.  At a temperature of
order $f_a$, a phase transition occurs in which the $U_{PQ}(1)$ symmetry
becomes spontaneously broken.  This is called the PQ phase transition.
At these temperatures, the non-perturbative QCD effects which produce the
effective potential $V(\overline\theta)$ are negligible \cite{GPY}, the
axion is massless and all values of $\langle a(x)\rangle$ are equally
likely.  Axion strings appear as topological defects.  One must
distinguish two scenarios, depending on wether inflation occurs 
with reheat temperature lower (case 1) or higher (case 2) than the 
PQ transition temperature.  In case 1 the axion field gets homogenized 
by inflation and the axion strings are 'blown' away.

When the temperature approaches the QCD scale, the potential
$V(\overline\theta)$ turns on and the axion acquires mass.  There is
a critical time, defined by $m_a(t_1)t_1 = 1$, when the axion field
starts to oscillate in response to the turn-on of the axion mass.
The corresponding temperature $T_1 \simeq 1$ GeV \cite{ac}.  The
initial amplitude of this oscillation is how far from zero the axion 
field lies when the axion mass turns on.  The axion field oscillations 
do not dissipate into other forms of energy and hence contribute to the
cosmological energy density today \cite{ac}. This contribution is called
of `vacuum realignment'.  It is further described below.  Note that the
vacuum realignment contribution may be accidentally suppressed in case 1
because the homogenized axion field happens to lie close to zero.

In case 2 the axion strings radiate axions \cite{rd,Har} from the
time of the PQ transition till $t_1$ when the axion mass turns on.   At
$t_1$ each string becomes the boundary of $N$ domain walls.  If $N=1$,
the network of walls bounded by strings is unstable \cite{Ev,Paris} and
decays away.  If $N>1$ there is a domain wall problem \cite{adw} because
axion domain walls end up dominating the energy density, resulting in a
universe very different from the one observed today.  There is a way
to avoid this problem by introducing an interaction which slightly
lowers one of the $N$ vacua with respect to the others.  In that
case, the lowest vacuum takes over after some time and the domain walls
disappear.  There is little room in parameter space for that to happen
and we will not consider this possibility further here.  A detailed
discussion is given in Ref.~\cite{axwall}.  Henceforth, we assume $N=1$.

In case 2 there are three contributions to the axion cosmological
energy density.  One contribution \cite{rd,Har,thA,Hag,Shel,Yam,us}
is from axions that were radiated by axion strings before $t_1$.  A
second contribution is from axions that were produced in the decay
of walls bounded by strings after $t_1$ \cite{Hag,Ly,Nag,axwall}.  A
third contribution is from vacuum realignment \cite{ac}.

Let me briefly indicate how the vacuum alignment contribution is
evaluated.  Before time $t_1$, the axion field did not oscillate
even once.  Soon after $t_1$, the axion mass is assumed to change
sufficiently slowly that the total number of axions in the
oscillations of the axion field is an adiabatic invariant.  The
number density of axions at time $t_1$ is
\begin{equation}
n_a(t_1)\simeq {1\over 2} m_a(t_1) \langle a^2(t_1)\rangle \simeq
\pi f_a^2 {1\over t_1}
\label{nat1}
\end{equation}
In Eq.~(\ref{nat1}), we used the fact that the axion field $a(x)$
is approximately homogeneous on the horizon scale $t_1$.  Wiggles
in $a(x)$ which entered the horizon long before $t_1$ have been
red-shifted away \cite{Vil}.  We also used the fact that the initial
departure of $a(x)$ from the nearest minimum is of order $f_a$.  The
axions of Eq.~(\ref{nat1}) are decoupled and non-relativistic.
Assuming that the ratio of the axion number density to the 
entropy density is constant from time $t_1$ till today, one 
finds \cite{ac,axwall}
\begin{equation}
\Omega_a \simeq {1 \over 2}
\left({0.6~10^{-5}\hbox{\ eV}\over m_a}\right)^{7\over 6}
\left({0.7 \over h}\right)^2
\label{oma}
\end{equation}
for the ratio of the axion energy density to the critical density
for closing the universe.  $h$ is the present Hubble rate in units
of 100 km/s.Mpc.  The requirement that axions do not overclose the
universe implies the constraint $m_a \gtwid 6 \cdot 10^{-6}$~ eV.

The contribution from axion string decay has been debated over the
years.  The main issue is the energy spectrum of axions radiated
by axion strings.  Battye and Shellard \cite{Shel} have carried out
computer simulations of bent strings (i.e. of wiggles on otherwise
straight strings) and have concluded that the contribution from
string decay is approximately ten times larger than that from vacuum
realignment, implying a bound on the axion mass approximately ten
times more severe, say $m_a \gtwid 6 \cdot 10^{-5}$ eV instead of
$m_a \gtwid 6 \cdot 10^{-6}$ eV.  My collaborators and I have done
simulations of bent strings \cite{Hag}, of circular string loops
\cite{Hag,us} and non-circular string loops \cite{us}.  We conclude
that the string decay contribution is of the same order of magnitude
than that from vacuum realignment.  Yamaguchi, Kawasaki and Yokoyama
\cite{Yam} have done computer simulations of a network of strings in 
anexpanding universe, and concluded that the contribution from string
decay is approximately three times that of vacuum realignment.  The
contribution from wall decay has been discussed in detail in 
ref.~\cite{axwall}.  It is probably subdominant compared to the 
vacuum realignment and string decay constributions.

It should be emphasized that there are many sources of uncertainty
in the cosmological axion energy density aside from the uncertainty
about the contribution from string decay.  The axion energy density may
be diluted by the entropy release from heavy particles which decouple
before the QCD epoch but decay afterwards \cite{ST}, or by the entropy
release associated with a first order QCD phase transition.  On the other
hand, if the QCD phase transition is first order \cite{pt}, an abrupt
change of the axion mass at the transition may increase $\Omega_a$. 
If inflation occurs with reheat temperature less than $T_{PQ}$, there 
may be an accidental suppression of $\Omega_a$ because the homogenized 
axion field happens to lie close to a $CP$ conserving minimum.  Because 
the RHS of Eq.~(7) is multiplied in this case by a factor of order the 
square of the initial vacuum misalignment angle ${a(t_1)\over f_a}$ which
is randomly chosen between $-\pi$ and $+\pi$, the probability that
$\Omega_a$ is suppressed by a factor $x$ is of order $\sqrt{x}$.  This
rule cannot be extended to arbitrarily small $x$ however because quantum
mechanical fluctuations in the axion field during the epoch of inflation 
do not allow the suppression to be perfect \cite{inflax}.  Recently,
Kaplan and Zurek proposed a model \cite{KZ} in which the axion decay
constant $f_a$ is time-dependent, the value $f_a(t_1)$ during the QCD
phase-transition being much smaller than the value $f_a$ today.  
This yields a suppression of the axion cosmological energy density by 
a factor $({f_a(t_1) \over f_a})^2$ compared to the usual case
[replace $f_a$ by $f_a(t_1)$ in Eq.~(\ref{nat1})].

The axions produced when the axion mass turns on during the QCD phase
transition are cold dark matter (CDM) because they are non-relativistic
from the moment of their first appearance at 1~GeV temperature.  Studies
of large scale structure formation support the view that the dominant
fraction of dark matter is CDM.  Any form of CDM necessarily contributes
to galactic halos by falling into the gravitational wells of galaxies.
Hence, there is excellent motivation to look for axions as constituent
particles of our galactic halo.

Finally, let's mention that there is a particular kind of clumpiness
\cite{amc,axwall} which affects axion dark matter if there is no inflation
after the Peccei-Quinn phase transition.  This is due to the fact that the
dark matter axions are inhomogeneous with $\delta \rho / \rho \sim 1$ over
the horizon scale at temperature $T_1 \simeq$ 1 GeV, when they
are produced at the start of the QCD phase-transition, combined
with the fact that their velocities are so small that they do not
erase these inhomogeneities by free-streaming before the time $t_{eq}$
of equality between the matter and radiation energy densities when
matter perturbations can start to grow.  These particular inhomogeneities
in the axion dark matter are in the non-linear regime immediately after
time $t_{eq}$ and thus form clumps, called `axion mini-clusters'
\cite{amc}.  They have mass $M_{mc} \simeq 10^{-13} M_\odot$ and
size $l_{mc} \simeq 10^{13}$ cm.

\section{Dark matter axion detection}

An electromagnetic cavity permeated by a strong static magnetic field
can be used to detect galactic halo axions \cite{ps}.  The relevant
coupling is given in Eq.~(\ref{aEB}). Galactic halo axions have 
velocities $\beta$ of order $10^{-3}$ and hence their energies
$E_a=m_a+{1\over 2} m_a\beta^2$ have a spread of order $10^{-6}$
above the axion mass.  When the frequency $\omega=2\pi f$ of a
cavity mode equals $m_a$, galactic halo axions convert resonantly
into quanta of excitation (photons) of that cavity mode.  The power
from axion $\to$ photon conversion on resonance is found to
be \cite{ps,kal}:
\begin{eqnarray}
P=\left ({\alpha\over\pi} {g_\gamma\over f_a}\right )^2 V\, B_0^2
\rho_a C {1\over m_a} \hbox{Min}(Q_L,Q_a)~~~~\nonumber\\
= 0.5\; 10^{-26} \hbox{Watt}\left( {V\over 500\hbox{\ liter}}\right)
\left({B_0\over 7\hbox{\ Tesla}}\right)^2 \nonumber\\
\cdot~C \left({g_\gamma \over 0.36}\right)^2
\left({\rho_a\over {1\over 2} \cdot 10^{-24}
{{\rm gr} \over \hbox{\rm cm}^3}}\right) \nonumber\\
\cdot~\left({m_a\over 2\pi (\hbox{GHz})}\right)\hbox{Min}(Q_L,Q_a)
\end{eqnarray}
where $V$ is the volume of the cavity, $B_0$ is the magnetic field
strength, $Q_L$ is its loaded quality factor, $Q_a=10^6$ is the
`quality factor' of the galactic halo axion signal (i.e. the ratio of
their energy to their energy spread), $\rho_a$ is the density of
galactic halo axions on Earth, and $C$ is a mode dependent form factor
given by
\begin{equation}
C = {\left| \int_V d^3 x \vec E_\omega \cdot \vec B_0\right|^2
\over B_0^2 V \int_V d^3x \epsilon |\vec E_\omega|^2}  \,
\end{equation}
where $\vec B_0(\vec x)$ is the static magnetic field,
$\vec E_\omega(\vec x) e^{i\omega t}$ is the oscillating electric
field and $\epsilon$ is the dielectric constant.  Because the axion 
mass is only known in order of magnitude at best, the cavity must be
tunable and a large range of frequencies must be explored seeking a
signal.  The cavity can be tuned by moving a dielectric rod or metal 
post inside it.  

For a cylindrical cavity and a homogeneous longitudinal magnetic field,
$C=0.69$ for the lowest TM mode.  The form factors of the other modes 
are much smaller.  The resonant frequency of the lowest TM mode of a
cylindrical cavity is $f$=115 MHz $\left( {1m\over R}\right)$ where $R$ 
is the radius of the cavity.  Since $10^{-6}\hbox{\ eV} = 2\pi$ (242 MHz),
a large cylindrical cavity is convenient for searching the low frequency
end of the range of interest.  To extend the search to high frequencies
without sacrifice in volume, one may power-combine many identical cavities
which fill up the available volume inside a magnet's bore \cite{rsci,hag}.  
This method allows one to maintain $C=0(1)$ at high frequencies, albeit
at the cost of increasing engineering complexity as the number of
cavities increases.

\begin{figure}
  \includegraphics[height=.2\textheight]{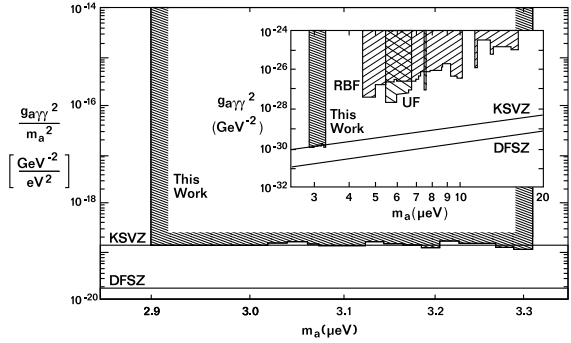}
  \caption{Axion couplings and masses excluded by the ADMX
experiment.  Also shown are the KSVZ and DFSZ model predictions.  
Indicated on the insert are the regions excluded by the pilot 
experiments at Brookhaven National Laboratory (RBF) and the
University of Florida (UF).  All limits are scaled to the 
local halo density $\rho_a = 7.5~10^{-25}~$g/cm$^3$ \cite{Gates}.}
\label{ADMXlim}
\end{figure}

Axion dark matter searches were carried out at Brookhaven National
Laboratory \cite{RBF}, the University of Florida \cite{UF}, Kyoto
University \cite{Kyoto}, and by the ADMX collaboration 
\cite{PRL,PRD,ApJL,PRDRC,Duf} at Lawrence Livermore National 
Laboratory.  The limits published by the ADMX collaboration in 
ref. \cite{PRD} are reproduced in Fig.~\ref{ADMXlim}.  By definition, 
$g_{a\gamma\gamma} \equiv {\alpha g_\gamma \over \pi f_a}$. Updated 
limits can be found in refs. \cite{ApJL,PRDRC}.

The ADMX experiment is equipped with a high resolution spectrometer 
which allows us to look for narrow peaks in the spectrum of microwave
photons caused by discrete flows, or streams, of dark matter axions in 
our neighborhood. In many discussions of cold dark matter detection
it is assumed that the distribution of CDM particles in galactic halos 
is isothermal. However, there are excellent reasons to believe that a
large fraction of the local density of cold dark matter particles is 
in discrete flows with definite velocities \cite{is}.  Indeed, because 
CDM has very low primordial velocity dispersion and negligible 
interactions other than gravity, the CDM particles lie on a 3-dim. 
hypersurface in 6-dim.  phase-space.  This implies that the velocity
spectrum of CDM particles at any physical location is discrete, i.e., 
it is the sum of distinct flows each with its own density and velocity.  

We searched for the peaks in the spectrum of microwave photons from 
axion to photon conversion that such discrete flows would cause in 
the ADMX detector.  We found none and placed limits \cite{Duf} on 
the density  of any local flow of axions as a function of the flow 
velocity dispersion over the axion mass range 1.98 to 2.17 $\mu$eV.
Our limit on the density of discrete flows is approximately a factor 
three more severe than our limit on the total local axion dark matter
density.

The ADMX experiment is presently being upgraded to replace the 
HEMT (high electron mobility transistors) receivers we have used 
so far with SQUID microwave amplifiers. HEMT receivers have noise 
temperature $T_n \sim 3~K$ \cite{Bradley} whereas $T_n \sim 0.05~K$ 
was achieved with SQUIDs \cite{Clarke}.  In a second phase of the 
upgrade, the experiment will be equipped with a dilution refrigerator 
to take full advantage of the lowered electronic noise temperature.  
When both phases of the upgrade are completed, the ADMX detector will 
have sufficient sensitivity to detect DFSZ axions at even a fraction 
of the local halo density.

\section{Solar axion detection}

The conversion of axions to photons in a magnetic field can also be 
used to look for solar axions \cite{ps,KVB,Laz,Min,Zio}.  The flux 
of solar axions on Earth is   ${7.4\cdot 10^{11} \over
{\rm sec}~{\rm cm}^2} ({g_\gamma \over 0.36})^2 ({m_a \over {\rm eV}})^2$
from the Primakoff conversion of thermal photons in the Sun \cite{KVB}.
The actual flux may be larger because other processes, such as
Compton-like scattering, contribute if the axion has an appreciable
coupling to the electron.  At any rate the flux is huge compared to 
what can be produced by man-made processes on Earth and it is cost
free.  Solar axions have a broad spectrum of energies of order the
temperature in the solar core, from one to a few keV.

Since the magnetic field is homogeneous on the length scale set by the
axion de Broglie wavelength, the final photon is colinear with the initial
axion.  The photon and axion also have the same energy assuming the
magnetic field is time-independent.  The $a \rightarrow \gamma$ conversion
probability is \cite{ps,KVB}
\begin{equation}
p = {1 \over 4} (g_{a\gamma\gamma})^2 (B_{0\perp} L F(q))^2
\label{prob}
\end{equation}
if $B_{0\perp} (z) = B_{0\perp} b(z)$ is the magnetic field transverse
to the direction of the colinear axion and photon, $z$ is the coordinate
along this direction, $L$ is the depth over which the magnetic field
extends and $F(q)$ is the form factor
\begin{equation}
F(q) = {1 \over L} \mid \int_0^L dz e^{iqz} b(z) \mid
\end{equation}
where $q = k_\gamma - k_a = E_a - \sqrt{E_a^2 - m_a^2}
\simeq {m_a^2 \over 2E_a}$ is the momentum transfer.  If the magnetic
field is homogeneous ($b=1$), then
\begin{eqnarray}
F(q) &=& {2 \over qL} \mid sin{qL \over 2} \mid \nonumber\\
&\simeq& 1 ~~~~~~~~{\rm for}~~~~qL \ll 1~~~\ .
\label{ff}
\end{eqnarray}
For $qL \gg 1$, the conversion probability goes as $sin^2({qL \over 2})$
because the axion and photons oscillate into each other back and forth.
The form factor $F(q)$ can be improved by filling the conversion region
with a gas whose pressure is adjusted in such a way that the plasma
frequency, which acts as an effective mass for the photon, equals the
axion mass \cite{KVB}.

\begin{figure}[t]
\includegraphics[height=.3\textheight]{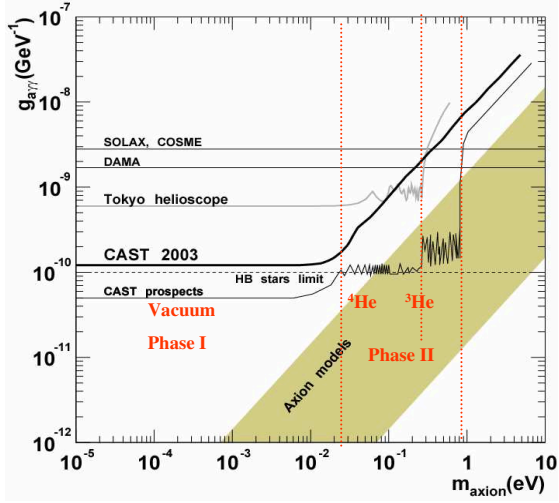}
\caption{Regions of $(m_a, g_{a\gamma\gamma})$ space 
ruled out by various solar axion searches.}
\label{CAST}
\end{figure}

Multiplying the flux times the conversion probability, one obtains the
event rate:
\begin{equation}
{{\rm events} \over {\rm time}} \simeq
{200 \over {\rm day}} ~{V L \over {\rm meter}^4} ~F(q)^2
({B_{0\perp} \over 8 {\rm Tesla}})^2 ({m_a \over {\rm eV}})^4 ~\ .
\end{equation}
The final state photons are soft x-rays which may be detected with good
efficiency.  There are radioactive backgrounds to worry about, however.

The above type of detector is usually referred to as an axion helioscope.
If a signal is found due to axions or familons, the detector becomes a 
marvelous tool for the study of the solar interior. Experiments were
carried out at Brookhaven National Lab. \cite{Laz}, the University of
Tokyo \cite{Min}, and more recently at CERN by the CAST collaboration
\cite{Zio}.  Fig.~\ref{CAST} shows the recently published CAST limits, 
as well as previous limits from the Tokyo group, and from the DAMA
\cite{DAMA} and SOLAX \cite{Avi} collaborations.  The latter two
experiments searched for solar axions by looking for their Primakoff 
conversion to photons in NaI and Ge crystals respectively.

\section{Laser experiments}

Eqs.(\ref{prob},\ref{ff}) give the conversion probability in a 
static magnetic field of an axion to a photon of the same energy. 
The polarization of the photon is parallel to the component of the
magnetic field transverse to the direction of motion.  The inverse
process, conversion of such a photon to an axion, occurs with the
same probability $p$.  K. van Bibber et al. \cite{shine} proposed 
a 'shining light through walls' experiment in which a laser beam 
is passed through a long dipole magnet like those used for 
high-energy physics accelerators.  In the field of the magnet, 
a few of the photons convert to axions.  Another dipole magnet is set
up in line with the first, behind a wall.  Since the axions go through
the wall unimpeded, this setup allows one to 'shine light through
the wall.'  An experiment of this type was carried out by the RBF
collaboration \cite{Ruoso}.  Also a proposal has been made to use
decommissioned HERA dipole magnets \cite{Ringwald} for this purpose.
Compared with a solar axion search, the 'shining light through walls' 
experiment has the advantage of greater control over experimental
parameters.  But the signal is much smaller because one pays twice 
the price of the very small axion-photon conversion rate.

It was also proposed \cite{Maiani} to look for the effect of the axion
on the propagation of light through a magnetic field.  If the photon 
beam is linearly polarized and the polarization direction is at an 
angle to the direction of the magnetic field, the plane of polarization
turns because the polarization component parallel to the magnetic
field gets depleted whereas the perpendicular component does not.  There
is an additional effect of birefringence because the component of light
polarized parallel to the magnetic field mixes with the axion and hence
moves more slowly than in vacuo.  Birefringence affects the ellipticity 
of the polarization as the light travels on.  Experiments searching for 
polarization rotation and birefringence of laser light passing through 
a magnetic field were carried out by the the RBF \cite{Cameron} and PVLAS 
\cite{Zav} collaborations.  In both experiments, the optical path was 
boosted by passing the laser beam many times through an optical cavity 
within the magnet. A positive result has very recently been reported 
by the PVLAS collaboration.
 
The PVLAS collaboration \cite{Zav} finds an optical rotation of 
$(3.9 \pm 0.5)~10^{-12}$ radians per pass, at 5 Tesla.  When 
interpreted in terms of an axion-like particle $a$ with 
coupling to two photons ${\cal L}_{a\gamma\gamma} = 
{- 1 \over M_a} a \vec{E} \cdot \vec{B}$, the range 
of mass and coupling consistent with the PVLAS effect is
$0.7~{\rm meV} \ltwid m_a \ltwid 2$ meV and  
$1 \cdot 10^5 {\rm GeV} \ltwid M_a \ltwid 6 \cdot 10^5$ GeV. 
At face value the PVLAS result disagres with the limits on 
axion-like particles from stellar evolution and solar axion 
searches, since ${1\over M_a}$ is the same as $g_{a\gamma\gamma}$ 
in Fig.~\ref{CAST}.  On the other hand, there may be ways to 
make the various results consistent with one another, as 
proposed for instance in ref.~\cite{Mas}.


\begin{thebibliography}{9}

\bibitem{abj}
S. Adler, Phys. Rev. {\bf 177} (1969) 2426; J.S. Bell and R. Jackiw,
Nuov. Cim. {\bf 60A} (1969) 47.

\bibitem{SW}
S. Weinberg, Phys. Rev. {\bf D11} (1975) 3583.

\bibitem{ned}
I.S. Altarev et al., Phys. Lett. {\bf B276} (1992) 242;
K.F. Smith et al., Phys. Lett. {\bf B234} (1990) 191.

\bibitem{KM}
M. Kobayashi and K. Maskawa, Progr. Theor. Phys. {\bf 49} (1973) 652.

\bibitem{PQ}
R. D. Peccei and H. Quinn, Phys. Rev. Lett. {\bf 38} (1977) 1440 and Phys.
Rev. {\bf D16} (1977) 1791.

\bibitem{WW}
S. Weinberg, Phys. Rev. Lett. {\bf 40} (1978) 223; F. Wilczek, Phys.
Rev. Lett. {\bf 40} (1978) 279.

\bibitem{VW}
C. Vafa and E. Witten, Phys. Rev. Lett. {\bf 53} (1984) 535.

\bibitem{GR}
H. Georgi and L. Randall, Nucl. Phys. {\bf B276} (1986) 241.

\bibitem{KSVZ}
J. Kim, Phys. Rev. Lett. {\bf 43} (1979) 103; M. A. Shifman, A. I.
Vainshtein and V. I. Zakharov, Nucl. Phys. {\bf B166} (1980) 493.

\bibitem{DFSZ}
M. Dine, W. Fischler and M. Srednicki, Phys. Lett. {\bf B104} (1981)
199; A. P. Zhitnitskii, Sov. J. Nucl. {\bf 31} (1980) 260.

\bibitem{Kim}
J.E. Kim, Phys. Rev. {\bf D31} (1985) 1733.

\bibitem{curr}
S. Weinberg in ref.\cite{WW};
W.A. Bardeen and S.-H.H. Tye, Phys. Lett. {\bf B74} (1978) 229;
J. Ellis and M.K. Gaillard, Nucl. Phys. {\bf B150} (1979) 141;
T.W. Donnelly et al., Phys. Rev. {\bf D18} (1978) 1607;
D.B. Kaplan, Nucl. Phys. {\bf B260} (1985) 215;
M. Srednicki, Nucl. Phys. {\bf B260} (1985) 689;
P. Sikivie, in 'Cosmology and Particle Physics', ed. E. Alvarez et al.,
World Scientific, 1987, pp 143-169.

\bibitem{arev}
Axion reviews include: J.E. Kim, Phys. Rep. {\bf 150} (1987) 1;
H.-Y. Cheng, Phys. Rep. {\bf 158} (1988) 1; R.D. Peccei, in
'CP Violation', ed. by C. Jarlskog, World Scientific Publ., 1989,
pp 503-551; M.S. Turner, Phys. Rep. {\bf 197} (1990) 67;
G.G. Raffelt, Phys. Rep. {\bf 198} (1990) 1.

\bibitem{Dicus}
D. Dicus, E. Kolb, V. Teplitz and R. Wagoner, Phys. Rev. {\bf D18}
(1978) 1829 and Phys. Rev. {\bf D22} (1980) 839.

\bibitem{Raff87}
G. Raffelt and D. Dearborn, Phys. Rev. {\bf D36} (1987) 2211.

\bibitem{Schramm}
D. Dearborn, D. Schramm and G. Steigman, Phys. Rev. Lett. {\bf 56}
(1986) 26.

\bibitem{1987a}
J. Ellis and K. Olive, Phys. Lett. {\bf B193} (1987) 525;
G. Raffelt and D. Seckel, Phys. Rev. Lett. {\bf 60} (1988) 1793;
M. Turner, Phys. Rev. Lett. {\bf 60} (1988) 1797;
H.-T. Janka et al., Phys. Rev. Lett. {\bf 76} (1996) 2621;
W. Keil et al.. Phys. Rev. {\bf D56} (1997) 2419.

\bibitem{GPY}
D. J. Gross, R. D. Pisarski and L. G. Yaffe, Rev. Mod. Phys. {\bf 53}
(1981) 43.

\bibitem{ac}
L. Abbott and P. Sikivie, Phys. Lett. {\bf B120} (1983) 133;
J. Preskill, M. Wise and F. Wilczek, Phys. Lett. {\bf B120} (1983) 127;
M. Dine and W. Fischler, Phys. Lett. {\bf B120} (1983) 137.

\bibitem{rd}
R. Davis, Phys. Rev. {\bf D32} (1985) 3172 and Phys. Lett. {\bf B180}

\bibitem{Har}
D. Harari and P. Sikivie, Phys. Lett. {\bf B195} (1987) 361.

\bibitem{Ev}
A. Vilenkin and A.E. Everett, Phys. Rev. Lett. {\bf 48} (1982) 1867.

\bibitem{Paris} P. Sikivie  in {\it Where are the elementary particles},
Proc. of the 14th Summer School on Particle Physics, Gif-sur-Yvette, 1982,
edited by P. Fayet et al. (Inst. Nat. Phys. Nucl. Phys. Particules, Paris,
1983).

\bibitem{adw}
P. Sikivie, Phys. Rev. Lett. {\bf 48} (1982) 1156.

\bibitem{axwall}
S. Chang, C. Hagmann and P. Sikivie, Phys. Rev. {\bf D59} (1999) 023505.

\bibitem{thA}
A. Vilenkin and T. Vachaspati, Phys. Rev. {\bf D35} (1987) 1138;
R.L. Davis and E.P.S. Shellard, Nucl. Phys. {\bf B324} (1989) 167;
A. Dabholkar and J. Quashnock, Nucl. Phys. {\bf B333} (1990) 815.

\bibitem{Hag}
C. Hagmann and P. Sikivie, Nucl. Phys. {\bf B363} (1991) 247.

\bibitem{Shel}
R.A. Battye and E.P.S. Shellard, Nucl. Phys. {\bf B423} (1994) 260,
Phys. Rev. Lett. {\bf 73} (1994) 2954 and erratum-ibid. {\bf 76}
(1996) 2203.

\bibitem{Yam}
M. Yamaguchi, M. Kawasaki and J. Yokoyama, Phys. Rev. Lett. {\bf 82}
(1999)
4578.

\bibitem{us}
C. Hagmann, S. Chang and P. Sikivie, Phys. Rev. {\bf D63} (2001) 125018.

\bibitem{Ly}
D. Lyth, Phys. Lett. {\bf B275} (1992) 279;

\bibitem{Nag}
M. Nagaswa and M. Kawasaki, Phys. Rev. {\bf D50} (1994) 4821.

\bibitem{Vil}
A. Vilenkin, Phys. Rev. Lett. {\bf 48} (1982) 59.

\bibitem{ST}
P. J. Steinhardt and M. S. Turner, Phys. Lett. {\bf B129} (1983) 51;
G. Lazarides, R. Schaefer, D. Seckel and Q. Shafi, Nucl. Phys. {\bf B346}
(1990) 193.

\bibitem{pt}
W. G. Unruh and R. M. Wald, Phys. Rev. {\bf D32} (1985) 831; M. S.
Turner, Phys. Rev. {\bf D32} (1985) 843; T. DeGrand, T. W. Kephart and
T. J. Weiler, Phys. Rev. {\bf D33} (1986) 910; M. Hindmarsh, Phys. Rev.
{\bf D45} (1992) 1130.

\bibitem{inflax}
A. D. Linde, JETP Lett. {\bf 40} (1984) 1333 and Phys. Lett. {\bf B158}
(1985) 375; D. Seckel and M. Turner, Phys. Rev. {\bf D32} (1985) 3178;
D. H. Lyth, Phys. Lett. {\bf B236} (1990) 408; A. D. Linde and D. H.
Lyth, Phys. Lett. {\bf B246} (1990) 353; M. Turner and F. Wiczek, Phys.
Rev. Lett. {\bf 66} (1991) 5; A. Linde, Phys. Lett. {\bf B259} (1991) 38;
D.H. Lyth, Phys. Rev. {\bf D45} (1992) 3394; D.H. Lyth and E.D. Stewart,
Phys. Lett. {\bf B283} (1992) 189 and Phys. Rev. {\bf D46} (1992) 532.

\bibitem{KZ}
D.B. Kaplan and K.M. Zurek, hep-ph/0507236.

\bibitem{amc}
C. J. Hogan and M. J. Rees, Phys. Lett. {\bf B205} (1988) 228;
E. Kolb and I. I. Tkachev, Phys. Rev. Lett. {\bf 71} (1993) 3051,
Phys. Rev. {\bf D49} (1994) 5040, and Ap. J. {\bf 460} (1996) L25.

\bibitem{ps}
P. Sikivie, Phys. Rev. Lett. {\bf 51} (1983) 1415 and Phys.
Rev. {\bf D32} (1985) 2988.

\bibitem{kal}
L. Krauss, J. Moody, F. Wilczek and D. Morris, Phys. Rev. Lett.
{\bf 55} (1985) 1797.

\bibitem{rsci}
C. Hagmann et al., Rev. Sci. Inst. {\bf 61} (1990) 1076.

\bibitem{hag}
C. Hagmann, Ph. D. thesis, unpublished.

\bibitem{RBF}
S. DePanfilis et al., Phys. Rev. Lett. {\bf 59} (1987) 839 and
Phys. Rev. {\bf 40} (1989) 3153.

\bibitem{UF}
C. Hagmann et al., Phys. Rev. {\bf D42} (1990) 1297.

\bibitem{Kyoto}
S. Matsuki and K. Yamamoto, Phys. Lett. {\bf B263} (1991) 523;
S. Matsuki, I. Ogawa, K. Yamamoto, Phys. Lett. {\bf B336} (1994) 573;
I. Ogawa, S. Matsuki and K. Yamamoto, Phys. Rev. {\bf D53} (1996) R1740;
A. Kitagawa, K. Yamamoto and S. Matsuki, hep-ph/9908445.

\bibitem{PRL}
C. Hagmann et al., Phys. Rev. Lett. {\bf 80} (1998) 2043.

\bibitem{PRD}
S.J. Asztalos et al., Phys. Rev. {\bf D64} (2001) 092003.

\bibitem{ApJL}
S.J. Asztalos et al., Ap. J. Lett. {\bf 571} (2002) L27.

\bibitem{PRDRC}
S.J. Asztalos et al., Phys. Rev. {\bf D69} (2004) 011101 (R).

\bibitem{Duf}
L. Duffy et al., Phys. Rev. Lett. {\bf 95} (2005) 091304.

\bibitem{Gates}
E.I. Gates, G. Gyuk and M.S. Turner, Astroph. J. {\bf 449} (1995) 123.

\bibitem{is}
J.R. Ipser and P. Sikivie, Phys. Lett. {\bf B291} (1992) 288.

\bibitem{Bradley}
R.F. Bradley, Nucl. Phys. B Proc. Suppl. {\bf 72} (1999) 137.

\bibitem{Clarke}
M. M{\" u}ck et al., Appl. Phys. Lett. {\bf 72} (1998) 2885;
Nucl. Phys. B Proc. Suppl. {\bf 72} (1999) 145.

\bibitem{KVB}
K. van Bibber, P. McIntyre, D. Morris and G. Raffelt, Phys. Rev.
{\bf D39} (1989) 2089.

\bibitem{Laz}
D. Lazarus et al., Phys. Rev. Lett. {\bf 69} (1992) 2089.

\bibitem{Min}
S. Moriyama et al., Phys. Lett. {\bf B434} (1998) 147.

\bibitem{Zio}
K. Zioutas et al., Phys. Rev. Lett. {\bf 94} (2005) 121301.

\bibitem{DAMA}
R. Bernabei et al., Phys. Lett. {\bf B515} (2001) 6.

\bibitem{Avi}
F.T. Avignone et al., Phys. Rev. Lett. {\bf 81} (1998) 5068;
R.J. Creswick et al., Phys. Lett. {\bf B427} (1998) 235.

\bibitem{shine}
K. van Bibber et al., Phys. Rev. Lett. {\bf 59} (1987) 759.

\bibitem{Ruoso}
G. Ruoso et al., Z. Phys. {\bf C56} (1992) 505.

\bibitem{Ringwald}
A. Ringwald, Phys.Lett.{\bf B569} (2003) 51.

\bibitem{Maiani}
L. Maiani, R. Petronzio and E. Zavattini, Phys. Lett. {\bf B175}
(1986) 359; G. Raffelt and L. Stodolsky, Phys. Rev. {\bf D37}
(1988) 1237.

\bibitem{Cameron}
Y. Semertzidis et al., Phys. Rev. Lett. {\bf 64} (1990) 2988;
R. Cameron et al., Phys. Rev. {\bf D47} (1993) 3707.

\bibitem{Zav}
E. Zavattini et al., hep-ex/0507107.

\bibitem{Mas}
E. Masso and J. Redondo, hep-ph/0504202.

\end{thebibliography}
\end{document}